\documentclass[sigconf]{acmart}

\usepackage{makecell}

\setcopyright{none}
\renewcommand\footnotetextcopyrightpermission[1]{}

\begin{document}

\title{RankEvolve: Automating the Discovery of Retrieval Algorithms via LLM-Driven Evolution}

\author{Jinming Nian}
\affiliation{%
  \institution{Santa Clara University}
  \city{Santa Clara}
  \state{CA}
  \country{USA}}
\email{jnian@scu.edu}
\author{Fangchen Li}
\affiliation{%
  \institution{Independent Researcher}
  \city{Bothell}
  \state{WA}
  \country{USA}}
\email{fangchen.li@outlook.com}
\author{Dae Hoon Park}
\affiliation{%
  \institution{Walmart Global Tech}
  \city{Sunnyvale}
  \state{CA}
  \country{USA}}
\email{dae.hoon.park@walmart.com}
\author{Yi Fang}
\affiliation{%
  \institution{Santa Clara University}
  \city{Santa Clara}
  \state{CA}
  \country{USA}}
\email{yfang@scu.edu}

\renewcommand{\shortauthors}{Nian et al.}

\begin{abstract}
Retrieval algorithms like BM25 and query likelihood with Dirichlet smoothing remain strong and efficient first-stage rankers, yet improvements have mostly relied on parameter tuning and human intuition. We investigate whether a large language model, guided by an evaluator and evolutionary search, can automatically discover improved lexical retrieval algorithms. We introduce RankEvolve, a program evolution setup based on AlphaEvolve, in which candidate ranking algorithms are represented as executable code and iteratively mutated, recombined, and selected based on retrieval performance across 12 IR datasets from BEIR and BRIGHT. RankEvolve starts from two seed programs: BM25 and query likelihood with Dirichlet smoothing. The evolved algorithms are novel, effective, and show promising transfer to the full BEIR and BRIGHT benchmarks as well as TREC DL 19 and 20. Our results suggest that evaluator-guided LLM program evolution is a practical path towards automatic discovery of novel ranking algorithms. Code is available \href{https://github.com/fangchenli/ranking-evolved}{\textcolor{blue}{\underline{here}}}.

\end{abstract}




\maketitle

\begin{figure} [t] 
  \centering
  \includegraphics[width=\linewidth]{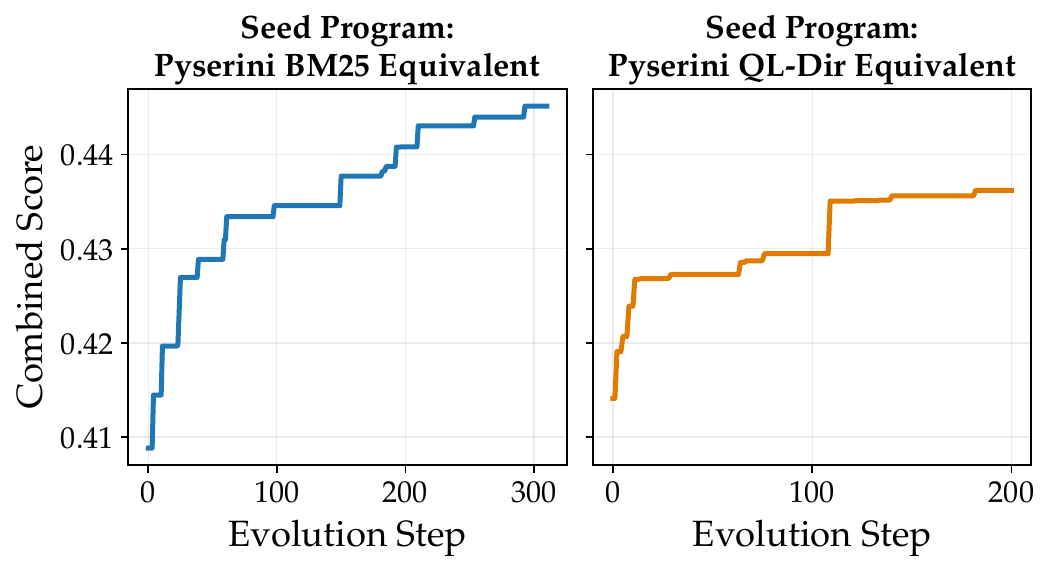}
  \caption{Combined score over evolution steps for two seed programs. The combined score is the optimization target, defined as $0.8 \times \text{Avg Recall@100} + 0.2 \times \text{Avg nDCG@10}$, averaged across 12 IR datasets\protect\footnotemark.} 
  \label{fig:evolve_plot}
\end{figure}
\footnotetext{BEIR: ArguAna, FiQA, NFCorpus, SciFact, SciDocs, TREC-COVID. BRIGHT: Biology, Earth Science, Economics, Pony, StackOverflow, TheoremQA.}

\section{Introduction}
Lexical retrieval algorithms such as BM25~\cite{DBLP:journals/ftir/RobertsonZ09} and query-likelihood (QL)~\cite{original_ql} remain remarkably strong and efficient first-stage rankers. Over the decades, numerous variants have been proposed, including BM25T~\cite{BM25T}, BM25-adapt~\cite{BM25_adapt}, BM25+~\cite{BM25_plus}, BM25L~\cite{BM25L}, and alternative QL smoothing methods such as Dirichlet smoothing and Jelinek–Mercer~\cite{ql_smoothing}. Yet these improvements have largely relied on parameter tuning and human intuition over individual scoring components. This raises a natural question: can we automate the discovery of improved lexical retrieval algorithms? Recent developments in large language models (LLMs) for automated scientific discovery offer a promising path for such automation. Systems like AlphaEvolve~\cite{novikov2025alphaevolvecodingagentscientific}, Evolution of Heuristics~\cite{EoH}, and ShinkaEvolve~\cite{lange2025shinkaevolveopenendedsampleefficientprogram} represent candidate solutions as executable code and leverage LLMs to iteratively mutate and recombine programs, with performance-driven selection. These methods have produced strong results in mathematics and combinatorial optimization, but have not yet been applied to information retrieval. We introduce RankEvolve, a program evolution setup in which each candidate ranking algorithm is a self-contained Python program ($\sim$300 lines) scored by an evaluator across multiple datasets. The evaluator's per-dataset metrics, a single fitness score, sampled top-performing and inspirational candidates, and prior attempted changes together form the prompt for a coding LLM to propose the next mutation. Based on AlphaEvolve's description, the candidates are managed by a combination of MAP-Elites~\cite{MAP_elites} and island-based evolutionary databases~\cite{island}. 

Starting from two seed programs, BM25 and query likelihood with Dirichlet smoothing, we evolve for several hundred steps. The resulting algorithms are novel and effective, introducing scoring mechanisms absent from either seed family. To assess generalization, we evaluate the evolved programs on held-out BEIR~\cite{BEIR} and BRIGHT~\cite{BRIGHT} subsets, as well as TREC DL 2019~\cite{DL19}, and DL 2020~\cite{DL20}, and find promising transfer beyond the datasets used during evolution. Our contributions are: (1) RankEvolve is the first attempt at evolving entire retrieval algorithms via LLM-guided program search; (2) we study the importance of seed program design, examining how structural freedom and abstraction affect the evolution outcome; and (3) we show that RankEvolve can discover algorithms with novel scoring motifs that transfer to unseen datasets, suggesting that evaluator-guided program evolution with LLMs is a promising way forward for automating IR research. 

\section{Related Work}
The idea of automatically discovering ranking functions has been explored through genetic programming (GP). \citet{arranger} proposed ARRANGER, a framework that uses GP to discover ranking functions from arithmetic operators ($+$, $\times$, $\log$) applied to IR features (tf, idf, dl). \citet{evolve_term_weighting} evolved local and global term-weighting schemes using a similar GP setup, producing functions that competed with BM25. A simpler alternative is grid search over existing hyperparameters~\cite{grid_search}, though this is limited to tuning a fixed function. RankEvolve differs in a fundamental way: classical GP evolves expression trees of arithmetic primitives by randomly swapping subtrees without understanding what the expression computes. RankEvolve uses an LLM as the mutation operator over a Python program, enabling reason-informed edits (e.g., recognizing that a signal for the document's coverage of the query is missing and introducing one) and a far richer search space. The learning-to-rank (LTR) paradigm~\cite{LTR} is also related, though the key distinction is that LTR combines existing features through learned weights, while RankEvolve proposes entirely new features and also evolves the relevance scoring function itself. 

\begin{table*}[t]
\setlength{\tabcolsep}{8pt} 
  \centering
  \caption{Macro-averaged results on unseen BRIGHT (6), BEIR (8), and TREC DL 19/20 datasets. Best per-group scores \textbf{bolded}. $^\star$Evolved with RankEvolve. $^\dagger$Significant over the respective seed (per-query paired $t$-test, $p<0.05$).}
  \label{tab:main_results}
  {%
  \begin{tabular}{lcccccc}
    \toprule
    & \multicolumn{2}{c}{BRIGHT} & \multicolumn{2}{c}{BEIR} & \multicolumn{2}{c}{TREC DL} \\
    \cmidrule(lr){2-3} \cmidrule(lr){4-5} \cmidrule(lr){6-7}
    Method & nDCG@10 & R@100 & nDCG@10 & R@100 & nDCG@10 & R@100 \\
    \midrule
    BM25~\cite{DBLP:journals/ftir/RobertsonZ09}   & 10.55 & 32.11 & \textbf{48.16} & 70.95 & 62.16 & 46.02 \\
    BM25+~\cite{BM25_plus}                         &  9.72 & 31.54 & 45.71 & 69.93 & 61.07 & 46.69 \\
    BM25-adpt~\cite{BM25_adapt}                    & 11.05 & 34.67 & 44.62 & 69.55 & 60.90 & \textbf{48.23} \\
    BM25$^\star$                                    & \textbf{11.79}$^\dagger$ & \textbf{37.51}$^\dagger$ & 47.90 & \textbf{72.43}$^\dagger$ & \textbf{64.57}$^\dagger$ & 47.10 \\
    \midrule
    QL-Dir~\cite{ql_smoothing}                                  &  9.52 & 32.48 & 44.15 & 68.72 & 57.03 & 46.69 \\
    QL-JM~\cite{ql_smoothing}                                   & 10.17 & 31.08 & 40.61 & 65.79 & 56.65 & 41.05 \\
    QL-Dir$^\star$                                  & \textbf{11.42}$^\dagger$ & \textbf{36.33}$^\dagger$ & \textbf{46.46}$^\dagger$ & \textbf{70.22}$^\dagger$ & \textbf{60.68}$^\dagger$ & \textbf{47.96} \\
    \bottomrule
  \end{tabular}%
  }
\end{table*}

\section{Method}
\label{sec:method}
We frame the discovery of improved retrieval algorithms as program synthesis via LLM-guided evolutionary search. RankEvolve iteratively mutates a seed program using an LLM and selects among variants based on retrieval performance. The overall pipeline consists of four components: (1)~a seed program and a system prompt, which together define the search space; (2)~a population database maintaining diversity via island-based evolution~\cite{island} and MAP-Elites~\cite{MAP_elites}; (3)~mutation guided by structured prompts; and (4)~an evaluator that computes fitness over retrieval datasets. 

\subsection{Search Space}
\label{sec:search_space}
The seed program and system prompt jointly define the search space. The seed program defines the evolvable interface: code regions that the LLM is permitted to modify. To maximize evolutionary freedom, we decompose each ranking function into a small number of abstract components. For the BM25 seed, we define three evolvable components: document representation, query representation, and scoring function. The initial behavior reproduces classic BM25:
\begin{equation}
  \text{BM25}(q,d) = \sum_{t \in q} \text{IDF}(t) \cdot
  \frac{\text{tf}(t, d) \cdot (k_1 + 1)}{\text{tf}(t, d) + k_1 \cdot \left(1 - b + b
  \cdot \frac{|d|}{\text{avgdl}}\right)}
  \label{eq:bm25}
\end{equation}
\begin{equation}
  \text{IDF}(t) = \log \left(\frac{N - \text{df}(t) + 0.5}{\text{df}(t) + 0.5}\right), 
  \label{eq:idf}
\end{equation}
where $\text{tf}(t, d)$ is the term frequency of $t$ in document $d$, $\text{df}(t)$ is the number of documents that contain $t$, $|d|$ is the document length, $\text{avgdl}$ is the average document length, and $k_1=0.9$ and $b=0.4$ following Pyserini~\cite{pyserini} defaults. For the QL-Dir seed, we add a fourth component, the collection language model, to the evolvable interface. The initial program implements: 
\begin{equation}
  \text{QL-Dir}(q, d) = \sum_{t \in q} \log \left(
  \frac{\text{tf}(t, d) + \mu \cdot P(t \mid C)}{|d| + \mu}\right),
  \label{eq:qldir}
\end{equation}
where $P(t \mid C)=\text{tf}(t,C)/|C|$ is the collection language model probability of term $t$, $C$ is the corpus, and $\mu=2000$ following Pyserini. The system prompt guides how to evolve the seed program. It details the design principles that encourage exploration of information-theoretic, probabilistic, and fundamentally novel ideas while discouraging ad-hoc constraints without justification. It also specifies the optimization objective, metrics, datasets, and evolvable versus restricted components. Together, these elements aim to maximize the LLM's freedom while ensuring every candidate remains a valid, executable retrieval system. 

\subsection{Population Management}
\label{sec:population}
We maintain a population of candidate programs using a combination of island-based evolution~\cite{island} and MAP-Elites~\cite{MAP_elites}. For the island-based model, the population is partitioned into $K$ independently evolving islands. Each island maintains its own MAP-Elites grid, including a record of its best programs. Programs inherit their parent's island, preserving lineage isolation. Within each island, MAP-Elites maps programs to a grid defined by two dimensions: complexity (code length) and diversity (edit distance from the population). Each dimension is divided into $B$ bins, yielding $B\times B$ cells per island. A new program is accepted into a cell only if the cell is unoccupied or the new program has strictly higher optimization target score than the current occupant. This mechanism is crucial for avoiding local minima, because always picking the best-performing candidate for mutation would quickly stagnate the search. Every $M$ iterations, the top $\gamma$ fraction of programs from each island migrate to adjacent islands to encourage new variants. Programs that have migrated previously are excluded to prevent duplication. 

\subsection{Mutation Proposal}
\label{sec:mutation}
At each iteration a parent program is selected from the current island via one of three strategies, chosen at random: (1) exploration (probability $p_e$): uniform random sampling from the island; (2) exploitation (probability $p_x$): sampling from the elite archive; or (3) weighted (probability $1-p_e -p_x$): performance-proportional sampling from the island. Additionally, we include $T$ best programs and $S$ randomly sampled programs from the same island to provide the LLM with diverse reference points. Each program is paired with its detailed evaluation metrics described in Section~\ref{sec:evaluator}. These, together with the system prompt, are presented to the LLM, which proposes a mutation in a \texttt{SEARCH/REPLACE} diff format. 

\subsection{Evaluator}
\label{sec:evaluator}
The evaluator imports a candidate program, executes its full pipeline (tokenization, indexing, and retrieval) on a set of evaluation datasets, and returns per-dataset nDCG@10, Recall@100, and latency measurements, all of which are stored alongside the program in the population database. For population management, sampling priority, and as the optimization target, we use a single fitness score: $0.8 \times \text{Avg Recall@100} + 0.2 \times \text{Avg nDCG@10}$, where the averages are taken across datasets. The weighting reflects the first-stage retrieval setting. The primary objective is to maximize recall of relevant documents for a downstream reranker, while nDCG@10 serves as a secondary signal for ranking quality of the retrieved set.

\section{Experiments}
\subsection{Baselines} BM25~\cite{DBLP:journals/ftir/RobertsonZ09} is the classical TF-IDF based ranking function with saturation and document length normalization. BM25+~\cite{BM25_plus} introduces a lower-bound term frequency normalization to address BM25's deficiency to overly penalize long documents, ensuring that additional occurrences of a query term always contribute positively to the score. BM25-adpt~\cite{BM25_adapt} extends this further by using document-adaptive $k_1$ parameters instead of a single global value. For language model baselines, QL-Dir~\cite{ql_smoothing} scores documents via query likelihood with Dirichlet smoothing (Section~\ref{sec:search_space}), while QL-JM~\cite{ql_smoothing} applies Jelinek-Mercer smoothing, where the query likelihood $P(q \mid d) = \sum_{t\in q}(1-\alpha)\, P(t \mid d) + \alpha\, P(t \mid C)$ is a linear combination between the maximum-likelihood document language model and the collection language model.

\subsection{Setup}
\label{sec:setup}
RankEvolve builds on OpenEvolve~\cite{openevolve}, an open-source implementation of AlphaEvolve. We use GPT-5.2 via API with temperature 0.85, and set $K=3$, $B=12$, $M=20$, $\gamma=0.15$, $T=4$, and $S=4$ (Sections~\ref{sec:population} and~\ref{sec:mutation}). RankEvolve ran for 300 steps from the BM25 seed and 200 steps from the QL-Dir seed, yielding the best-scoring programs at steps 293 and 182 respectively. Figure~\ref{fig:evolve_plot} shows the optimization target trajectory over the course of evolution. 
We evaluate on three benchmark suites: 14 datasets from BEIR~\cite{BEIR}, all 12 subsets of BRIGHT~\cite{BRIGHT}, and TREC Deep Learning 2019~\cite{DL19}/2020~\cite{DL20}, totaling 28 datasets. We exclude BioASQ, Signal-1M (RT), TREC-NEWS, and Robust04 as they are not publicly available on BEIR GitHub, and omit MSMARCO as it serves as the corpus for TREC DL. Of these 28 datasets, only 12 are used during evolution (from BEIR: ArguAna, FiQA, NFCorpus, SciFact, SciDocs, TREC-COVID; from BRIGHT: Biology, Earth Science, Economics, Pony, StackOverflow, TheoremQA). The remaining 16 are held out entirely and used solely to test generalization. 

\subsection{Results}
Table~\ref{tab:main_results} presents the macro-averaged results for each benchmark on the 16 held-out datasets. BM25$^\star$ outperforms all BM25 baselines on BRIGHT and BEIR Recall@100 and achieves the highest nDCG@10 on TREC DL. QL-Dir$^\star$ consistently outperforms both QL-Dir and QL-JM across all three benchmarks. Gains are statistically significant over the respective seeds on most evaluation groups, and extend to datasets not seen during evolution, indicating that RankEvolve discovers better retrieval algorithms that generalize well rather than overfitting to the evaluator signal. 

\begin{figure*} [t] 
  \centering
  \includegraphics[width=\linewidth]{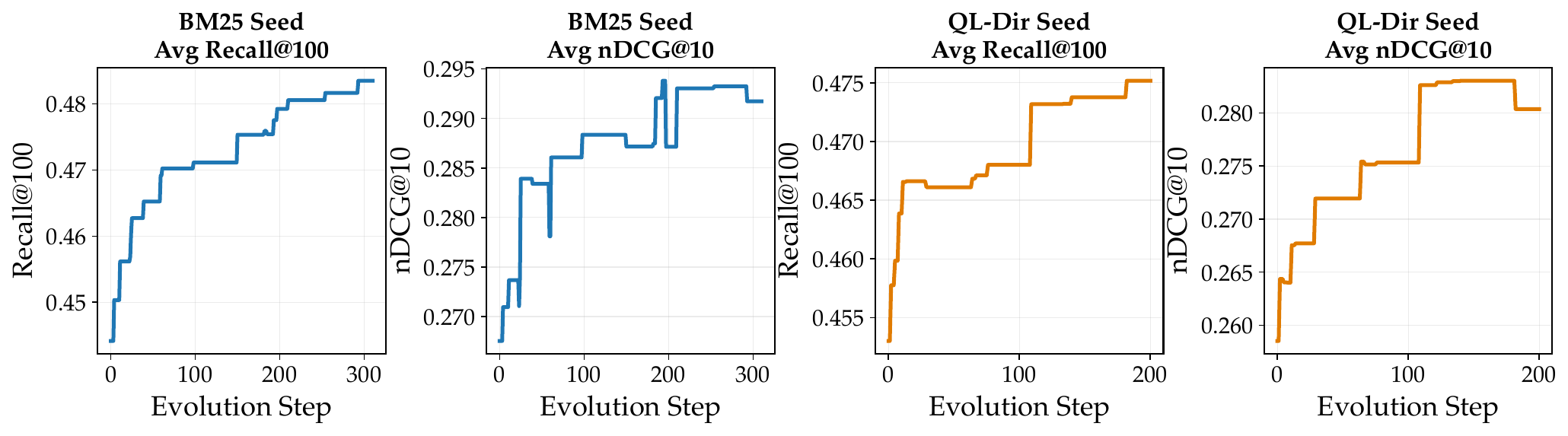}
  \caption{Evolution trajectories for the BM25 seed (left) and Dirichlet seed (right). Recall@100 improves nearly monotonically in both runs, while nDCG@10 occasionally regresses at the same steps, reflecting deliberate trades made by the evolutionary process to maximize the optimization target.} 
  \label{fig:4_panel}
\end{figure*}

Figure~\ref{fig:4_panel} shows the Recall@100 and nDCG@10 trajectories across both evolutionary runs. Recall@100 improves nearly monotonically, while nDCG@10 occasionally regresses. This is expected: the optimization target weights Recall ($0.8\times$) far more heavily than nDCG ($0.2\times$), so RankEvolve will accept any mutation that trades a small nDCG loss for a larger Recall gain. The pattern is visible in both seeds, with nDCG dipping at exactly the steps where Recall makes its largest jumps. The combined score~\ref{fig:evolve_plot} is monotonically increasing throughout optimization. When we optimize a weighted sum, we implicitly authorize the optimizer to sacrifice the pawn to advance the queen. RankEvolve simply makes this trade visible. 

To understand what RankEvolve discovered, we now analyze the best-evolved programs in detail.

\subsection{The Evolved BM25 Algorithm}
After 293 evolution steps, the best evolved algorithm converges to a multi-channel, modulated scoring function that operates entirely over lexical features, yet has substantially departed from BM25 in structure. The top-level scoring function is:
\begin{equation}
\label{eq:score_evolved_bm25}
\begin{split}
S(q,d) = \; & R(q^{\text{base}}, d^{\text{base}}) + w_{\text{pfx}} \cdot R(q^{\text{pfx}}, d^{\text{pfx}}) \\
             & + \; w_{\text{bi}} \cdot R(q^{\text{bi}}, d^{\text{bi}}) + w_{\text{mic}} \cdot G(q) \cdot R(q^{\text{mic}}, d^{\text{mic}}),
\end{split}
\end{equation}
where $R$ is a shared core scoring function applied across four parallel token spaces. Base tokenization uses the standard Lucene tokenizer. Prefix tokenization ($w_{\text{pfx}}=0.1$) truncates each token to its first 5 characters, acting as a cheap stemming approximation. Bigram tokenization ($w_{\text{bi}}=0.08$) concatenates consecutive token pairs. Micro tokenization ($w_{\text{mic}}=0.12$) uses rolling character 3-grams for sub-word matching, gated by $G(q) = \sigma((\overline{\text{IDF}}(q) - 2.2)/1.0)$, a sigmoid of the query's mean IDF that activates sub-word matching only for rare or technical queries.

The shared scoring function $R$ combines a base evidence term with a chain of bounded multipliers:
\begin{equation}
\label{eq:R}
R(q,d) = \frac{\ln(1 + E) \cdot B_{\text{cov}} \cdot B_{\text{spec}} \cdot B_{\text{coord}} \cdot B_{\text{anc}}}{B_{\text{len}}},
\end{equation}
where $E = \sum_{t \in M} w(t) \cdot \ln(1 + \text{tf}(t, d))$ accumulates weighted log-TF evidence over matched query terms $M$, and $B_*$ are multipliers of the form $1 + \gamma \cdot (\cdot)$ with small $\gamma$ that modulate the score based on query-term coverage, topical specificity, term coordination, rare-term anchoring, and document length. Their individual effects are gentle, but their combined effect can be substantial. The outer $\ln(1+E)$ applies a second layer of saturation, making the scoring robust to outlier term frequencies through double log-compression. Full definitions of all components appear in Appendix~\ref{app:evolved_bm25}.

We highlight three aspects of the evolved design that are especially notable. First, the composite term weight $w(t)$ multiplies three separate functions of $\text{IDF}(t)$, which together suppress stopword-like terms (low IDF) while leaving rare terms nearly unaffected. \emph{RankEvolve has learned a soft stopword filter without ever being told about stopwords.} Second, the specificity multiplier $B_{\text{spec}}$ uses pointwise mutual information between each query term and the candidate document to reward documents where matched terms appear with higher-than-expected frequency, an idea closely related to the collection language model in probabilistic retrieval. Third, the length dampener $B_{\text{len}} = 1 + 0.15 \cdot \ln(1 + (|d|+1)/(\text{avgdl}+1))$ replaces BM25's linear normalization with a gentler logarithmic form, aligning with findings from \citet{BM25_plus} that BM25 over-penalizes long documents.

RankEvolve arrived at all of these components through evolutionary search alone, without any explicit guidance toward them. The prefix channel approximates stemming, the multi-layered IDF weighting acts as a soft stopword filter, the PMI-based multiplier ($B_{\text{spec}}$) resembles a collection language model, and the logarithmic length normalization independently addresses a known BM25 weakness. To the best of our knowledge, no exact formulation in the existing literature matches the evolved scoring function, yet these well-studied concepts emerged naturally from a purely metric-driven search process. This suggests that such ideas may have been abundant enough in the LLM's training data to leave a strong impression in its parametric knowledge, or that they are perhaps inevitable solutions to the lexical retrieval problem, or both. 

\begin{table*}[t]
\centering
\caption{Ablation on code structure design (BM25 family). Each row reports the best program found by RankEvolve under a given level of structural freedom; evolve step and total search budget are shown in parentheses. The optimization target is $0.8 \times \text{R@100} + 0.2 \times \text{nDCG@10}$, macro-averaged over the 12 datasets used during evolution; unseen evaluation covers 16 held-out datasets (see Section~\ref{sec:setup}).}
\label{tab:code_structure_ablation}
{%
\begin{tabular}{l cc cc cc cc}
\toprule
& \multicolumn{2}{c}{\textbf{BRIGHT}} & \multicolumn{2}{c}{\textbf{BEIR}} & \multicolumn{2}{c}{\textbf{TREC DL}} & \multicolumn{2}{c}{\textbf{Optimization Target}} \\
\cmidrule(lr){2-3} \cmidrule(lr){4-5} \cmidrule(lr){6-7} \cmidrule(lr){8-9}
\textbf{Code Structure} & nDCG@10 & R@100 & nDCG@10 & R@100 & nDCG@10 & R@100 
  & Seen (12 datasets) & Unseen (16 datasets) \\
\midrule
Original (step 0)               & 10.55 & 32.11 & 48.16 & 70.95 & 62.16 & 46.02 & 40.30 & 49.77 \\
Constrained (step 115/200)      & 11.36 & 34.68 & \textbf{48.27} & 70.75 & 61.86 & 45.44 & 42.87 & 50.47 \\
Composable (step 185/200)       & \textbf{11.97} & 37.27 & 46.95 & 70.30 & 63.07 & \textbf{48.62} & 43.31 & 51.33 \\
Freeform (step 177/200)         & 11.39 & 36.92 & 46.87 & 71.27 & 61.17 & 45.40 & 44.35 & 51.20 \\
Freeform (step 293/300)         & 11.79 & \textbf{37.51} & 47.90 & \textbf{72.43} & \textbf{64.57} & 47.10 & \textbf{44.58} & \textbf{52.22} \\
\bottomrule
\end{tabular}%
}
\end{table*}

\subsection{The Evolved Query Likelihood Algorithm}
After 182 evolution steps starting from classical query likelihood with Dirichlet smoothing, the best evolved algorithm retains the probabilistic language-modeling foundation but departs substantially from the standard formulation. The top-level scoring function is:
\begin{equation}
\label{eq:score_evolved_dir}
S(q,d) = \sum_{t \in q_u} \omega(t)\,\tilde{s}(t,d) \;+\; \sum_{t \in q_u} m(t,d) \;+\; \text{AND}(q,d) \;+\; \text{LP}(d),
\end{equation}
where $q_u$ is the set of unique query terms in the vocabulary. The score decomposes into four additive components: a weighted sum of per-term relevance scores $\tilde{s}$, a missing-term penalty $m$, a soft-AND coverage bonus, and a log-normal document length prior. Unlike the evolved BM25, which restructures scoring into a product of bounded multipliers (Equation~\ref{eq:R}), the evolved query likelihood algorithm retains an additive architecture, but augments it with coordination and penalty mechanisms that standard query likelihood lacks entirely. Full definitions of all components appear in Appendix~\ref{app:evolved_ql}.

We highlight four aspects that represent the most significant departures from the seed. First, RankEvolve replaces the standard collection language model $P(t \mid C)$ with a three-stage enriched estimate. The raw collection probabilities are raised to a power $\tau = 0.85$ and renormalized (flattening the distribution to give more mass to rare terms), then interpolated with a document-frequency language model $P_{\text{df}}(t) = \text{df}(t)/N$ that is robust to bursty documents, and finally mixed with a small uniform component as a safety floor. \emph{RankEvolve has independently discovered that flattening the collection language model improves retrieval}, related to the information-based retrieval models of \citet{ad_doc_ir}.

Second, raw term frequency is replaced by $\text{tf}(t,d)^{\beta(t)}$, where the exponent $\beta(t) \in [0.70, 1.0]$ is a function of normalized IDF. Common terms saturate aggressively ($\beta \approx 0.70$), while rare terms preserve the full TF signal ($\beta \approx 1.0$). Standard Dirichlet smoothing applies no TF saturation; BM25 applies uniform saturation; the evolved BM25 also uses uniform log-saturation. The per-term adaptive exponent is strictly more expressive than all of them.

Third, the evolved algorithm applies a leaky rectifier to per-term scores. Where standard implementations discard negative per-term contributions, the evolved function retains them at 12\% strength. Together with a separate missing-term penalty that applies the Dirichlet smoothing-only score (scaled to 7\%) for completely absent query terms, this creates a layered penalty architecture at two granularities: per-term weakness and per-term absence. The small magnitudes suggest the evolutionary process found that recall is fragile and penalties must be conservative.

Fourth, the document length prior $\text{LP}(d) = -0.06 \cdot (\log(|d|) - \log(\text{avgdl}))^2$ penalizes deviation in both directions from the corpus average on a log scale. This contrasts with the evolved BM25's length dampener, which only penalizes long documents. The quadratic form reflects the insight that very short documents are also problematic.

\subsection{Convergent Principles Across Seeds}
The evolved query likelihood algorithm retains a recognizable language-modeling structure: the core score is still a log-likelihood ratio with Dirichlet smoothing, and the missing-term penalty is literally the Dirichlet score at $\text{tf} = 0$. The evolutionary process chose to augment the probabilistic foundation rather than abandon it. This contrasts with the evolved BM25, which restructured the scoring function more radically into a product of bounded multipliers applied to a double-log-compressed evidence term. Despite starting from fundamentally different baselines, one from a probabilistic language model and the other from a TF-IDF based retrieval function, the two evolved algorithms exhibit convergence in their high-level strategies. Both independently discovered term-frequency saturation, soft stopword filtering, explicit coordination mechanisms, and gentle length normalization. Yet they implement these strategies through different architectural paradigms: multiplicative modulation versus additive penalties, multi-channel tokenization versus single-space gating, uniform log-saturation versus per-term adaptive exponents. The fact that both trajectories converged to the same abstract principles from very different starting points suggests that these are not artifacts of any particular scoring framework, but rather fundamental requirements of effective lexical retrieval.

\section{Ablation Study}
\paragraph{Effect of seed structure on search space and convergence.}
We study how the structural freedom of the seed program shapes evolution. We design three functionally identical BM25 seeds with increasing degrees of freedom, and evolve each using the procedure from Section~\ref{sec:method}. The ``constrained'' seed fixes the BM25 formula and permits only hyperparameter tuning and selection among predefined component variants (e.g., Lucene vs. Robertson IDF). This setting is analogous to automated grid search. The ``composable'' seed decomposes retrieval into modular primitives whose formulas may be rewritten or extended, but keeps the overall pipeline structure fixed. The ``freeform'' seed, used in our main experiments, defines only query representation, document representation, and scoring function. It fixes the evaluator interface and leaves everything else evolvable. 

Table~\ref{tab:code_structure_ablation} shows a clear trend: greater structural freedom yields monotonically higher optimization target scores on both the 12 datasets used by the evaluator and the 16 held-out datasets. Constrained evolution converges earliest but produces the smallest gains, confirming that parameter tuning alone has limited potential. The composable variant improves further by introducing novel scoring primitives, yet its fixed pipeline prevents deeper architectural changes. The freeform variant converges last but achieves the best scores on both seen and unseen datasets, demonstrating again that the improvements generalize rather than overfit to the development suite. These results suggest that seed program design sets an upper bound on what RankEvolve can discover. Restrictive seeds bias the search towards local optima near the original formulation, whereas seeds with more structural freedom expand the search space and allow evolution to identify non-obvious improvements. 

\paragraph{Complementary strengths across structures.}
Beyond the aggregate trend, Table~\ref{tab:code_structure_ablation} reveals that the best programs evolved from differently structured seeds exhibit complementary per-benchmark strengths. The freeform variant (step 293/300) achieves the highest scores on BRIGHT Recall@100, BEIR Recall@100, and TREC DL nDCG@10, yet it is not uniformly dominant: its TREC DL Recall@100 (47.10) lags behind the composable variant (48.62); its BEIR nDCG@10 (47.90) falls short of the constrained variant (48.27); and notably also under-perform the original BM25 baseline (48.16). 

These per-metric differences are not random noise. They reflect the distinct inductive biases each seed structure imposes on the search trajectory. The constrained seed, limited to selecting among predefined component variants and tuning their parameters, preserves the original BM25 formula almost intact. This conservatism prevents it from discovering novel scoring primitives, but it also shields against degradations on benchmarks where the classical formulation is almost near-optimal. The composable seed allows evolution to rewrite individual scoring primitives while keeping the pipeline skeleton fixed. This additional freedom lets it discover recall-oriented modifications (e.g., aggressive document-length normalization and alternative term-frequency saturation curves) that boost TREC DL Recall@100beyond what either the constrained or freeform variants achieve. The freeform seed removes even the pipeline constraint, enabling evolution to restructure the scoring architecture itself. This produces the largest aggregate gains but occasionally trades precision on narrow evaluation slices for broader improvements across the board. 

\paragraph{Influence on the discovery process.}
The fact that no single structural configuration dominates on every metric underscores a key insight: RankEvolve's value lies not only in the single best program it returns, but also in the diverse family of high-performing programs it is able to discover. Each seed structure guides evolution through a different region of program space, surfacing solutions that emphasize different trade-offs between nDCG and Recall, or between robustness on short web queries (TREC DL) and complex reasoning-intensive queries (BRIGHT). In practice, one could run RankEvolve from multiple seed structures and select or ensemble the best-performing variant per deployment scenario, treating the evolutionary trajectories as a form of structured exploration rather than a single-objective optimization. 

Despite these per-benchmark variations, the freeform seed remains the strongest choice on average: it achieves the highest optimization target on both the 12 seen datasets and the 16 unseen datasets, confirming that maximal structural freedom provides the best risk--award trade-off when a single general-purpose retrieval function is needed.  

\begin{table}[t]
  \centering
  \small
  \caption{Average per-document indexing latency and per-query retrieval latency across all 28 datasets. Lowest values are \textbf{bolded}, second lowest are \underline{underlined}. $^\star$Evolved with RankEvolve.}
  \label{tab:latency}
  \begin{tabular}{lcc}
    \toprule
    Method & \makecell{Indexing \\ (ms/doc)} & \makecell{Query \\ (ms/query)} \\
    \midrule
    BM25                              & \underline{1.79} & \textbf{56.72} \\
    BM25$^\star$ Constrained          & 1.85             & \underline{58.50} \\
    BM25$^\star$ Composable           & \textbf{1.77}    & 111.49 \\
    BM25$^\star$ Freeform (step 177)  & 2.37             & 171.52 \\
    BM25$^\star$ Freeform (step 293)  & 2.81             & 648.89 \\
    \midrule
    QL-Dir                            & \underline{2.02} & \textbf{178.26} \\
    QL-JM                             & \textbf{1.95}    & \underline{212.24} \\
    QL-Dir$^\star$ Freeform (step 182)& \underline{2.02}             & 325.41 \\
    \bottomrule
  \end{tabular}
\end{table}

\section{Latency} 
Table~\ref{tab:latency} reports per-document indexing and per-query retrieval latency averaged across all 28 datasets. Indexing overhead is negligible across all variants. Query latency increases with program complexity: the best performing BM25$^\star$ (step 293) is roughly $11\times$ slower than the seed BM25. This is entirely expected. While latency statistics are visible to the LLM during evolution, they are never part of the optimization target nor explicitly mentioned to be mindful of in the system prompt, so the search process has no pressure to favor efficient solutions. Incorporating latency as an explicit optimization objective is a straightforward extension that we leave to future work. 

Despite this, the latency profile offers a useful lens into how RankEvolve discovers improvements at different levels of structural freedom. The constrained variant adds virtually no overhead (58.50 vs. 56.72~ms/query), confirming that its gains stems almost entirely from parameter tuning rather than algorithmic breakthrough. Yet even this minimal-complexity evolution meaningfully improves the optimization target (Table~\ref{tab:code_structure_ablation}), showing that the simplest form of RankEvolve, which is effectively an automated parameter search, already provides value. 

Among the freeform variants, the trajectory from step~177 to step~293 is very interesting. At step~177 the program had already achieved strong effectiveness (Table~\ref{tab:code_structure_ablation}), with query latency still within a modest $3\times$ of the baseline. The following 116~steps of evolution continued to improve recall and nDCG, but at the cost of a $3.8\times$ further increase in query latency, indicating that later evolutionary steps exploit increasingly complex scoring mechanisms whose marginal effectiveness gains carry disproportionate computational cost. This mirrors a common pattern in program synthesis: early mutations tend to explore high-return structural changes, while later mutations add refinements that are effective but expensive. 

\section{Conclusion and Future Work}
We introduce RankEvolve, a framework that applies LLM-guided program evolution to discover lexical retrieval algorithms. Evolved from BM25 and QL-Dir seeds, the resulting functions consistently outperform their seeds and established variants on held-out benchmark datasets. The evolved programs independently rediscover and reformulate well-studied IR concepts. Our ablation confirms that the structural freedom of the seed program determines the ceiling of what RankEvolve can discover. The evolved algorithms, while effective, are much more complex than their seeds. Defining and encouraging elegance is a natural next step. More broadly, RankEvolve optimizes whatever objective the evaluator defines, which suggests the framework could extend beyond lexical retrieval to dense retrieval, learned sparse representations, and even LLM reranking methods. A straightforward extension is to incorporate efficiency constraints as an explicit optimization objective. We hope RankEvolve motivates further exploration of LLM-guided program evolution as a tool for automated IR research. 

\newpage

\bibliographystyle{ACM-Reference-Format}
\bibliography{reference}

\newpage
\appendix

\section{Full Evolved BM25 Scoring Function}
\label{app:evolved_bm25}

This appendix provides the complete definitions of all components in the evolved BM25 algorithm described in Section~4.4. The top-level scoring function (Equation~\ref{eq:score_evolved_bm25}) applies a shared core function $R$ across four parallel token spaces, and the core function $R$ (Equation~\ref{eq:R}) combines base evidence with a chain of bounded multipliers.

\paragraph{Term weight.}
The term weight $w(t)$ determines how much each query term can contribute to the evidence term $E$:
\begin{equation}
w(t) = \text{qtf}(t,q)^{0.5} \cdot \text{IDF}(t) \cdot \left(\frac{\text{IDF}(t)}{\text{IDF}(t) + 1}\right)^{0.6} \cdot \frac{\text{IDF}(t)}{\text{IDF}(t) + 1.25}
\end{equation}
where $\text{qtf}(t,q)$ is the query term frequency, and
\begin{equation}
\text{IDF}(t)=-\ln\left(\frac{\text{df}(t)+1}{N+2}\right),
\end{equation}
which is closer to the traditional TF-IDF formulation than to BM25's IDF. The first factor $\text{qtf}(t,q)^{0.5}$ gives diminishing credit to repeated query terms. The remaining three factors are all functions of $\text{IDF}$. Acting together, they behave like a stopword removal filter: a stopword-like term with $\text{IDF}\approx1$ is suppressed by all three factors and ends up with a very small weight, whereas a rare term with $\text{IDF}\approx8$ passes through nearly unaffected.

\paragraph{Coverage multiplier.}
The coverage multiplier rewards breadth of match:
\begin{equation}
B_{\text{cov}} = 1 + 0.25 \cdot \frac{W_M}{W},
\end{equation}
where $W_M = \sum_{t \in M} w(t)$ is the total weight of matched terms and $W = \sum_{t \in q} w(t)$ is the total weight of all query terms. The ratio $W_M/W \in [0,1]$ measures what fraction of the query's importance the document satisfies. A document matching all query terms receives a $1.25\times$ boost. BM25 has no explicit equivalent; it relies solely on additive accumulation to implicitly reward multi-term matches.

\paragraph{Specificity multiplier.}
The specificity multiplier rewards documents where matched terms appear with higher-than-expected frequency:
\begin{equation}
B_{\text{spec}} = 1 + 0.10 \cdot \frac{\sum_{t \in M^+} w(t) \cdot \min(\text{PMI}(t,d),\; 3.0)}{W},
\end{equation}
where $M^+ = \{t \in M : \text{PMI}(t,d) > 0\}$ and
\begin{equation}
\text{PMI}(t,d) = \ln\!\left(\frac{\text{tf}(t,d) \cdot N}{\max(|d|, 25) \cdot \text{df}(t)}\right)
\end{equation}
is the pointwise mutual information between term $t$ and document $d$. A positive PMI means the term is over-represented in the document relative to its corpus-wide rate. The cap at $3.0$ and the document length floor of $25$ prevent short documents from producing extreme values. $B_{\text{spec}}$ captures how strongly the document focuses on query terms.

\paragraph{Coordination multiplier.}
The coordination multiplier provides a second, calibrated signal for multi-term matching:
\begin{equation}
B_{\text{coord}} = 1 + 0.20 \cdot \frac{\tau_{\text{coord}}}{\tau_{\text{coord}} + \ln(1 + W)} \cdot \frac{|M|}{|q|},
\end{equation}
where $|M|$ is the number of matched query terms, $|q|$ is the number of unique query terms, and $\tau_{\text{coord}} = 2.5$. The calibration factor $\frac{\tau_{\text{coord}}}{\tau_{\text{coord}} + \ln(1+W)}$ adapts the bonus to query difficulty. For short, rare-term queries, $W$ is small and the calibration factor is close to $1$, so the bonus is modest because high per-term evidence already implicitly rewards coordination. For long, common-word queries, $W$ is large and the calibration factor shrinks, preventing double-counting with $B_{\text{cov}}$.

\paragraph{Anchor multiplier.}
The anchor multiplier is a recall safeguard for rare terms:
\begin{equation}
B_{\text{anc}} = 1 + 0.14 \cdot \ln(1 + A)
\end{equation}
\begin{equation}
A = \max_{t \in M:\, \text{IDF}(t) > 4.2} \frac{\text{IDF}(t) - 4.2}{\text{IDF}(t)}
\end{equation}
If a document matches even one very rare query term (IDF above $4.2$), it receives a small boost proportional to that term's rarity. Only the single rarest matched term contributes, keeping the boost bounded.

\paragraph{Length dampener.}
The length dampener replaces BM25's per-term linear normalization with a single global logarithmic penalty:
\begin{equation}
B_{\text{len}} = 1 + 0.15 \cdot \ln\!\left(1 + \frac{|d| + 1}{\text{avgdl} + 1}\right)
\end{equation}
This is gentler than BM25's $1 - b + b \cdot |d|/\text{avgdl}$. A document twice the average length is penalized by roughly $10\%$, compared to up to $40\%$ under BM25 with $b=0.4$. The logarithmic form means the penalty grows very slowly for extremely long documents, which benefits heterogeneous corpora with high length variance.

\section{Full Evolved Query Likelihood Scoring Function}
\label{app:evolved_ql}

This appendix provides the complete definitions of all components in the evolved query likelihood algorithm described in Section~4.5.

\paragraph{Enriched collection language model.}
The standard collection language model $P(t \mid C) = \text{tf}_C(t)/|C|$ is replaced by a three-stage estimate. First, the raw collection probabilities are raised to a power $\tau = 0.85$ and renormalized:
\begin{equation}
P_\tau(t) = \frac{P(t \mid C)^\tau}{\sum_{t' \in V} P(t' \mid C)^\tau},
\end{equation}
where $t'$ is a dummy variable indicating that the denominator sums across all terms in the vocabulary $V$. This flattens the distribution, transferring probability mass from frequent terms to rare ones. Second, the tempered model is interpolated with a document-presence language model $P_{\text{df}}(t) = \text{df}(t)/N$:
\begin{equation}
P_{\text{mix}}(t) = 0.90 \cdot P_\tau(t) + 0.10 \cdot P_{\text{df}}(t).
\end{equation}
The document-frequency model counts each document once regardless of internal repetition, providing a stabler estimate than the token-frequency model for terms concentrated in bursty documents. Third, a small uniform component is mixed in to ensure no term has zero background probability:
\begin{equation}
P_C(t) = 0.97 \cdot P_{\text{mix}}(t) + \frac{0.03}{|V|}.
\end{equation}

\paragraph{Adaptive term-frequency saturation.}
Raw term frequency is replaced by $\text{tf}_{\text{eff}}(t, d) = \text{tf}(t, d)^{\,\beta(t)}$, where
\begin{equation}
\beta(t) = 1 - 0.30 \cdot (1 - \text{IDF}_{01}(t))
\end{equation}
\begin{equation}
\text{IDF}_{01}(t) = \frac{\log\left(\frac{N+1}{\text{df}(t)+1}\right)}{\max_{t' \in V}\;\log\left(\frac{N+1}{\text{df}(t')+1}\right)}.
\end{equation}
The normalized IDF $\text{IDF}_{01}(t) \in [0,1]$ maps each term to its relative rarity. For common terms ($\text{IDF}_{01} \approx 0$), $\beta(t) \approx 0.70$ and term-frequency saturates aggressively. For rare terms ($\text{IDF}_{01} \approx 1$), $\beta(t) \approx 1.0$ and the full term-frequency signal is preserved.

\paragraph{Gated Dirichlet log-likelihood.}
The core per-term relevance score is:
\begin{equation}
s_{\text{base}}(t,d) = g(t) \cdot \log\!\left(\frac{1 + \frac{\text{tf}(t,d)^{\,\beta(t)}}{\mu \cdot P_C(t)}}{\frac{|d|+\mu}{\mu}}\right),
\end{equation}
with $\mu = 1750$, lower than the standard default of $\mu = 2000$ because the tempered background model already provides more uniform smoothing. The gate $g(t)$ is the Entity Dispersion Ratio (EDR):
\begin{equation}
g(t) = 1 + 0.45 \cdot \text{clip}\left(\log\left(\frac{P_{\text{df}}(t)}{P_C(t)}\right),\; -2.5,\; 2.5\right).
\end{equation}
The ratio $P_{\text{df}}(t)/P_C(t)$ measures whether a term's document spread matches its token frequency. Terms that appear across many documents but with low total frequency (high ratio) are upweighted as independently informative. Terms with high token frequency concentrated in few documents (low ratio) are downweighted. The clipping keeps the gate bounded in $[-0.125,\; 2.125]$.

\paragraph{Query term weight.}
Each query term receives a composite weight:
\begin{equation}
\omega(t) = \big(\text{qtf}(t) \cdot r(t)\big)^{0.6}
\end{equation}
\begin{equation}
r(t) = 1 + 0.9 \cdot \frac{\text{clip}\!\left(\log\left(\frac{P_{\text{df}}(t)}{P_C(t)}\right),\; 0,\; 2.5\right)}{2.5}
\end{equation}
The residual weight $r(t) \in [1.0, 1.9]$ boosts query terms that are spread broadly rather than concentrated in few documents. The outer power $0.6$ applies sub-linear damping to repeated query terms, preventing verbose or repetitive queries from over-counting.

\paragraph{Leaky rectifier.}
Per-term scores are passed through a leaky rectifier:
\begin{equation}
\tilde{s}(t, d) = \begin{cases} s_{\text{base}}(t, d) & \text{if } s_{\text{base}}(t, d) \geq 0, \\ 0.12 \cdot s_{\text{base}}(t, d) & \text{otherwise.} \end{cases}
\end{equation}
Negative evidence is retained at 12\% of its full strength, enough to break ties and mildly penalize documents that partially match but are weak on specific terms.

\paragraph{Missing-term penalty.}
An explicit penalty applies for each query term that is completely absent from the document:
\begin{equation}
m(t, d) = \mathbf{1}[\text{tf}(t,d) = 0] \cdot 0.07 \cdot \omega(t) \cdot \log\left(\frac{\mu \cdot P_C(t)}{|d| + \mu}\right).
\end{equation}
This is precisely the Dirichlet log-likelihood score evaluated at $\text{tf} = 0$, scaled down by $0.07$. Together with the leaky rectifier, this creates a layered penalty architecture at two granularities (per-term weakness and per-term absence).

\paragraph{Soft-AND coverage bonus.}
A bonus rewards documents matching more distinct query terms:
\begin{equation}
\text{AND}(q,d) = 0.14 \cdot \frac{1}{|q_u|} \sum_{t \in q_u} \tanh\frac{\omega(t) \cdot \max(\tilde{s}(t,d),\; 0)}{3.0}.
\end{equation}
The $\tanh$ approximates a binary ``this term was matched well'' signal. Summing and normalizing by $|q_u|$ gives a coverage fraction in $[0, 1]$. The $\tanh$ design is robust to the scale of per-term scores, so the coverage signal reflects purely breadth, not depth.

\paragraph{Document length prior.}
A Gaussian prior on log-length penalizes deviation from the corpus average in both directions:
\begin{equation}
\text{LP}(d) = -0.06 \cdot \big(\log(|d|) - \log(\text{avgdl}))^2.
\end{equation}
A document $10\times$ the average length receives a penalty of approximately $-0.32$, a nudge rather than a decisive factor.

\end{document}